\documentclass[aps,prb,floatfix,amsmath,amssymb,preprint,eqsecnum]{revtex4}
\usepackage{graphicx}
\usepackage{dcolumn}
\usepackage{bm}
\usepackage{epstopdf}
\usepackage{enumerate}
\usepackage{setspace}   
\usepackage{amsmath}
\usepackage{amssymb}

\begin{document}
\title{Finite temperature dissipation and transport\\
near quantum critical points}

\author{Subir Sachdev}
\affiliation{Department of Physics, Harvard University, Cambridge MA
02138}

\date{Oct 5, 2009\\
\vspace{1.6in}}
\begin{abstract}
I review a variety of model systems and their quantum critical points, motivated by recent experimental
and theoretical developments. These are used to present a general discussion of the non-zero temperature crossovers 
in the vicinity of a quantum critical point. Insights are drawn from the exact solutions of quantum critical transport
obtained from the AdS/CFT correspondence. I conclude with a discussion of 
the role of quantum criticality in the phase diagram of the cuprate
superconductors.\\
~\\
~\\
{\tt Contributed chapter to the book `Developments in Quantum Phase Transitions', edited by Lincoln Carr.}
\end{abstract}

\maketitle

The author's book \cite{ssbook} on quantum phase transitions has an extensive discussion on the 
dynamic and transport properties of a variety of systems at non-zero temperatures above
a zero temperature quantum critical point. The purpose of the present article is to briefly review some basic material, and 
to then update
the earlier discussion with a focus on experimental and theoretical developments in the decade
since the book was written. We note other recent reviews \cite{ssnaturephys} from which portions
of this article have been adapted.

We will begin in Section~\ref{sec:ssmodel} by introducing a variety of model systems and their
quantum critical points; these are motivated by recent experimental and theoretical developments.
We will use these systems to introduce basic ideas on the finite temperature crossovers near
quantum critical points in Section~\ref{sec:sscross}. In Section~\ref{sec:sstrans}, we will focus on the important {\em quantum critical
region\/} and present a general discussion of its transport properties.
An important recent development has been the complete exact solution of the dynamic and transport
properties in the quantum critical region of a variety of (supersymmetric) 
model systems in two and higher dimensions: this
will be described in Section~\ref{sec:ssexact}. The exact solutions are found to agree with the earlier
general ideas discussed here in Section~\ref{sec:sstrans}.
Quite remarkably, the exact solution proceeds via a mapping to the 
theory of black holes in one higher spatial dimension: we will only briefly mention this mapping here,
and refer the reader to the article by Hartnoll in this book for more information. 
As has often been the case in the history of physics, the existence of a new class of solvable models leads to new
and general insights which apply to a much wider class of systems, almost all 
of which are not exactly solvable. This has also been the case here, as we will review in Section~\ref{sec:sshydro}:
a hydrodynamic theory of the low frequency transport properties has been developed, and has led
to new relations between a variety of thermo-electric transport co-efficients. Finally, in Section~\ref{sec:sstc}
we will turn to the cuprate high temperature superconductors, and present recent proposals on how
ideas from the theory of quantum phase transition may help unravel 
the very complex phase diagram of these important materials.

\section{Model systems and their critical theories}
\label{sec:ssmodel}

\subsection{Coupled dimer antiferromagnets} 
\label{sec:sslgw}

\begin{figure}
\centering
 \includegraphics[width=\linewidth]{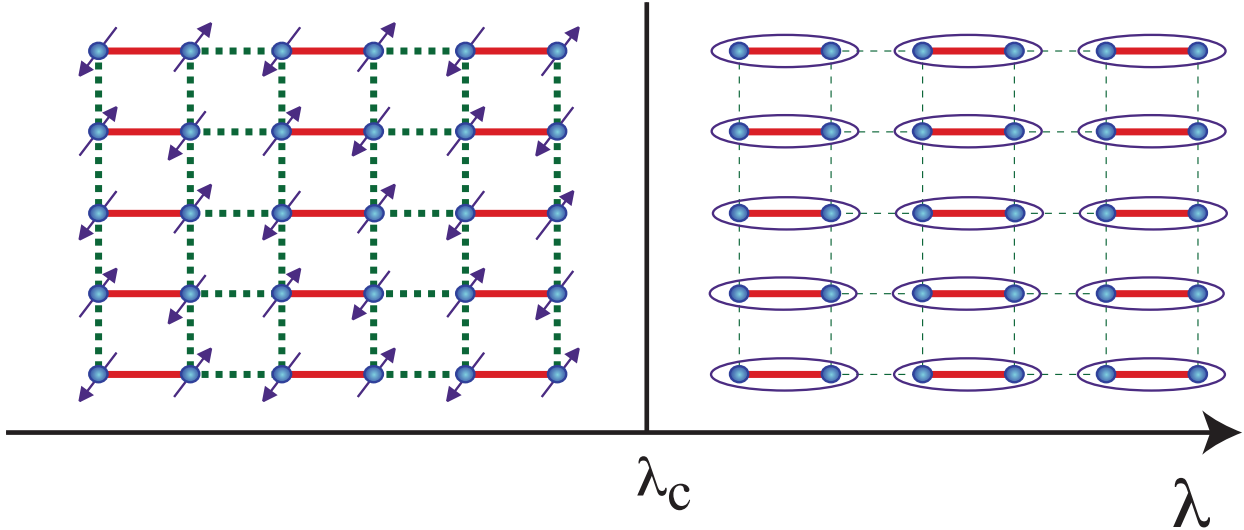}
 \caption{The coupled dimer antiferromagnet. The full red lines represent an exchange interaction $J$, while the dashed green lines have exchange $J/\lambda$. The ellispes represent a singlet valence
 bond of spins $(|\uparrow \downarrow \rangle - | \downarrow \uparrow \rangle )/\sqrt{2}$.}
\label{fig:ssdimer}
\end{figure}

Some of the best studied examples of quantum phase transitions arise in insulators with unpaired $S=1/2$ electronic
spins residing on the sites, $i$, of a regular lattice. Using $S^a_i$ ($a=x,y,z$) to represent the spin $S=1/2$ operator
on site $i$, the low energy spin excitations are described by the Heisenberg exchange Hamiltonian
\begin{equation}
H_J = \sum_{i<j} J_{ij} S^a_i \cdot S^a_j + \ldots
\label{eq:ssHJ}
\end{equation}
where $J_{ij} > 0$ is the antiferromagnetic exchange interaction. We will begin with a simple realization of 
this model is illustrated in Fig.~\ref{fig:ssdimer}. 
The $S=1/2$ spins reside on the sites of a square lattice, and have nearest neighbor exchange equal
to either $J$ or $J/\lambda$. Here $\lambda \geq 1$ is a tuning parameter which induces a quantum phase
transition in the ground state of this model. 

At $\lambda = 1$, the model has full square lattice symmetry,
and this case is known to have a N\'eel ground state which breaks spin rotation symmetry. This state has a
checkerboard polarization of the spins, just as found in the classical ground state, and as illustrated 
on the left side of Fig.~\ref{fig:ssdimer}. It can be characterized by a vector order parameter $\varphi^a$
which measures the staggered spin polarization
\begin{equation}
\varphi^a = \eta_i S^a_i
\end{equation}
where $\eta_i=\pm 1$ on the two sublattices of the square lattice. In the N\'eel state we have $\langle \varphi^a \rangle \neq  0$,
and we expect that the low energy excitations can be described by long wavelength fluctuations of a field $\varphi^a (x, \tau)$ over
space, $x$, and imaginary time $\tau$.

On the other hand, for $\lambda \gg 1$ it is evident from Fig.~\ref{fig:ssdimer} that the ground state preserves 
all symmetries of the Hamiltonian: it has total spin $S=0$ and can be considered to be a product of nearest
neighbor singlet valence bonds on the $J$ links. It is clear that this state cannot be smoothly connected
to the N\'eel state, and so there must at least one quantum phase transition as a function $\lambda$. 

Extensive quantum Monte Carlo simulations \cite{sstroyer,ssmatsu,ssjanke} 
on this model have shown there is a direct phase
transition between these states at a critical $\lambda_c$, as in Fig.~\ref{fig:ssdimer}. 
The value of $\lambda_c$
is known accurately, as are the critical exponents characterizing a second-order quantum phase
transition. These critical exponents are in excellent agreement with the simplest proposal for the critical
field theory, \cite{ssjanke} which can be obtained via conventional Landau-Ginzburg arguments. Given the vector
order parameter $\varphi^a$, we write down the action in $d$ spatial and one time dimension,
\begin{equation}
\mathcal{S}_{LG} = \int d^d r d\tau \left[ \frac{1}{2} \left[ (\partial_\tau \varphi^a )^2  + v^2 ( \nabla \varphi^a )^2 + s ( \varphi^a)^2 \right]
+ \frac{u}{4} \left[ (\varphi^a)^2 \right]^2 \right], \label{eq:ssslg}
\end{equation}
as the simplest action expanded in gradients and powers of $\varphi^a$ which is consistent will all
the symmetries of the lattice antiferromagnet.
The transition is now tuned by varying $s \sim (\lambda - \lambda_c)$. Notice that this model 
is identical to the Landau-Ginzburg theory for the thermal phase transition in a $d+1$ dimensional ferromagnet,
because time appears as just another dimension. As an example of the agreement: the critical exponent of the correlation
length, $\nu$, has the same value, $\nu = 0.711 \ldots$, to three significant digits in a quantum Monte Carlo study of the coupled
dimer antiferromagnet,\cite{ssjanke} and in a 5-loop analysis \cite{ssvicari} of the renormalization group fixed point of $\mathcal{S}_{LG}$
in $d=2$. 
Similar excellent agreement is obtained for the double-layer antiferromagnet \cite{sssandsca,ssmatsushita}
and the coupled-plaquette antiferromagnet.\cite{ssafa}

\begin{figure}
\centering
 \includegraphics[width=3.7in]{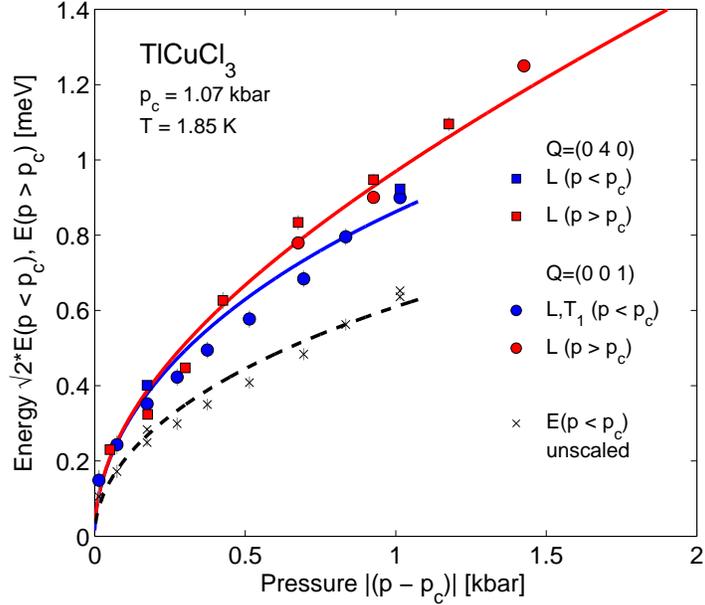}
 \caption{Energies of the gapped collective modes across the pressure ($p$) tuned quantum phase transition in
 TlCuCl$_3$ observed by Ruegg {\em et al.}\cite{ssruegg}. We test the description by the action $\mathcal{S}_{LG}$
 in Eq.~(\ref{eq:ssslg}) with $s \propto (p_c - p)$ by comparing $\sqrt{2}$ times the energy gap for $p<p_c$ with the
 energy of the longitudinal mode for $p>p_c$. The lines are the fits to a $\sqrt{|p-p_c|}$ dependence, testing the 1/2
 exponent.}
\label{fig:ssruegg}
\end{figure}

In experiments, the best studied realization of the coupled-dimer antiferromagnet is TlCuCl$_3$. In this crystal, the dimers are coupled
in all three spatial dimensions, and the transition from the dimerized state to the N\'eel state can be induced by application of pressure.
Neutron scattering experiments by Ruegg and collaborators \cite{ssruegg} have 
clearly observed the transformation in the excitation spectrum across the transition,
as is described by a simple fluctuations analysis about the mean field saddle point of $\mathcal{S}_{LG}$. In the dimerized phase
($s>0$), a triplet of gapped excitations is observed, corresponding to the three normal modes of $\varphi^a$ oscillating
about $\varphi^a = 0$; as expected, this triplet gap vanishes upon approaching the quantum critical point. In a mean field analysis,
valid for $d \geq 3$,
the field theory in Eq.~(\ref{eq:ssslg}) has a triplet gap of $\sqrt{s}$.
In the N\'eel phase, the neutron scattering detects 2 gapless spin waves, and one gapped longitudinal
mode \cite{ssnormand}. This is describe by $\mathcal{S}_{LG}$ for $s<0$, where $\varphi^a$ experiences
an inverted `Mexican hat' potential with a minumum at $|\varphi^a| = \sqrt{|s|/v}$. Expanding about this
minimum we find that in addition to the gapless spin waves, there is a mode involving amplitude
fluctuations of $|\varphi^a|$ which has an
energy gap of $\sqrt{2 |s|}$.
These mean field predictions for the energy of the gapped modes on the two sides of the transition are tested in Fig.~\ref{fig:ssruegg}:
the observations are in good agreement with the 1/2 exponent and the 
predicted \cite{sslg} $\sqrt{2}$ ratio, providing a non-trival experimental
test of the $\mathcal{S}_{LG}$ field theory. 

\subsection{Deconfined criticality}
\label{sec:ssdeconfine}

We now consider an analog of transition discussed in Section~\ref{sec:sslgw}, but for a Hamiltonian $H=H_0 + \lambda H_1$ which has
full square lattice symmetry at all $\lambda$. For $H_0$, we choose a form of $H_J$, with $J_{ij}= J$ for all nearest
neighbor links. Thus at $\lambda=0$ the ground state has N\'eel order, as in the left panel of Fig.~\ref{fig:ssdimer}.
We now want to choose $H_1$ so that increasing $\lambda$ leads to a spin singlet state with spin rotation symmetry restored.
A large number of choices have been made in the literature, and the resulting ground state invariably \cite{ssrsl} has valence bond solid
(VBS) order; a VBS state has been observed in the organic antiferromagnet EtMe$_3$P[Pd(dmit)$_2$]$_2$ \cite{sskato1,sskato2}.  
The VBS state is superficially similar to the dimer singlet state in the right panel of Fig.~\ref{fig:ssdimer}:
the spins primarily form valence bonds with near-neighbor sites. However, because of the square lattice symmetry of the Hamiltonian, a columnar arrangement of the valence bonds as in Fig.~\ref{fig:ssdimer}, breaks the square lattice rotation
symmetry; there are 4 equivalent columnar states, with the valence bond columns running along different directions. 
More generally, a VBS state is a spin singlet state, with a non-zero degeneracy due to a spontaneously broken lattice
symmetry. Thus a direct transition between the N\'eel and VBS states involves two distinct broken symmetries:
spin rotation symmetry, which is broken only in the N\'eel state, and a lattice rotation symmetry, which is broken only
in the VBS state. The rules of Landau-Ginzburg-Wilson theory imply that there can be no generic second-order
transition between such states.

It has been argued that a second-order N\'eel-VBS transition can indeed occur \cite{sssenthil}, but the critical theory is not expressed
directly in terms of either order parameter. It involves a fractionalized bosonic spinor $z_\alpha$ ($\alpha = \uparrow,
\downarrow$), and an emergent gauge field $A_\mu$. 
The key step is to express the vector field $\varphi^a$ in terms of $z_\alpha$ by
\begin{equation}
\varphi^a = z_\alpha^\ast {\sigma}^a_{\alpha \beta} z_\beta
\label{eq:ssPhiz}
\end{equation}
where ${\sigma}^a$ are the $2\times2$ Pauli matrices. Note that this mapping from $\varphi^a$ to $z_\alpha$
is redundant. We can make a spacetime-dependent change in the phase of the $z_\alpha$ by the field $\theta(x,\tau)$
\begin{equation}
z_\alpha \rightarrow e^{i \theta} z_\alpha
\label{eq:ssgauge}
\end{equation}
and leave $\varphi^a$ unchanged. All physical properties must therefore also be invariant under Eq.~(\ref{eq:ssgauge}),
and so the quantum field theory for $z_\alpha$ has a U(1) gauge invariance, much like that found in quantum electrodynamics.
The effective action for the $z_\alpha$ therefore requires introduction of an `emergent' 
U(1) gauge field $A_\mu$ (where $\mu = x, \tau$ is a 
three-component spacetime index). The field $A_\mu$ is unrelated the electromagnetic field, but is an internal
field which conveniently describes the couplings between the spin excitations of the antiferromagnet.  
As we did for $\mathcal{S}_{LG}$,
we can write down the quantum field theory for $z_\alpha$ and $A_\mu$ by the constraints of symmetry and gauge invariance,
which now yields
\begin{equation}
\mathcal{S}_z =  \int d^2 r d \tau \biggl[
|(\partial_\mu -
i A_{\mu}) z_\alpha |^2 + s |z_\alpha |^2  + u (|z_\alpha |^2)^2 + \frac{1}{2g^2}
(\epsilon_{\mu\nu\lambda}
\partial_\nu A_\lambda )^2 \biggl] \label{eq:ssSz}
\end{equation}
For brevity, we have now used a ``relativistically'' invariant notation, and scaled away the spin-wave velocity $v$; the values
of the couplings $s,u$ are different from, but related to, those in $\mathcal{S}_{LG}$. The Maxwell action for $A_\mu$ is generated from 
short distance $z_\alpha$ fluctuations, and it makes $A_\mu$ a dynamical field; its coupling $g$ is unrelated
to the electron charge. 
The action $\mathcal{S}_z$ is a valid description of the N\'eel state for $s<0$ (the critical upper value of $s$ will have fluctuation
corrections away from 0), where the gauge theory enters a Higgs phase with $\langle z_\alpha \rangle \neq 0$. This description of the N\'eel state as a Higgs phase has an analogy with the Weinberg-Salam theory of the weak interactions---in the latter case it is hypothesized that the condensation of a Higgs boson gives a mass to the $W$ and $Z$ gauge bosons, whereas here the condensation of $z_\alpha$ quenches the $A_\mu$ gauge boson.
As written, the $s>0$ phase of $\mathcal{S}_z$ is a `spin liquid' state with a $S=0$ collective gapless excitation associated with the 
$A_\mu$ photon. Non-perturbative effects \cite{ssrsl} associated with the monopoles in $A_\mu$ (not discussed here), show that this spin liquid is
ultimately unstable to the appearance of VBS order. 

Numerical studies of the N\'eel-VBS transition have focussed on a specific lattice antiferromagnet proposed by 
Sandvik \cite{sssandvik,sssandvik2,ssmelkokaul}. There is strong
evidence for VBS order proximate to the N\'eel state, along with persuasive evidence of a second-order transition.
However, some studies \cite{sswiese,sskuklov} support a very weak first order transition.

\subsection{Graphene}
\label{sec:ssgraphene}

The last few years have seen an explosion in experimental and theoretical studies \cite{ssneto} of graphene: a single hexagonal layer of carbon atoms.
At the currently observed temperatures, there is no evident broken symmetry in the electronic excitations, and so it is not
conventional to think of graphene as being in the vicinity of a quantum critical point. However, graphene does indeed undergo
a bona fide quantum phase transition, but one without any order parameters or broken symmetry. This transition may be viewed as 
being `topological' in character, and is associated with a change in nature of the Fermi surface as a function of carrier density.

\begin{figure}
\centering
 \includegraphics[width=3in]{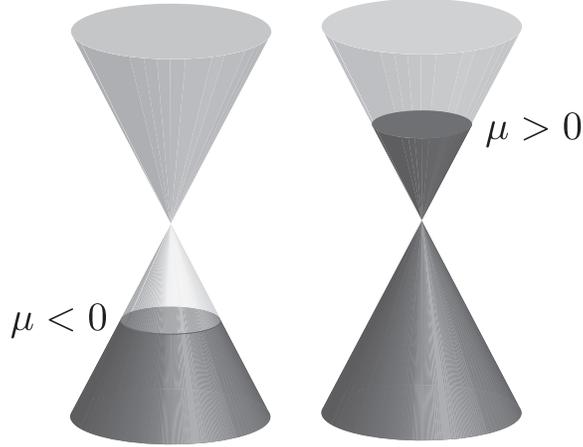}
 \caption{Dirac dispersion spectrum for graphene showing a `topological' quantum phase transition from a hole Fermi surface for $\mu<0$ to a electron Fermi surface
 for $\mu > 0$.}
\label{fig:ssgraphene}
\end{figure}

Pure, undoped graphene has a conical electronic dispersion spectrum at two points in the Brillouin zone, with the Fermi energy at the particle-hole
symmetric point at the apex of the cone. So there is no Fermi surface, just a Fermi point, where the electronic energy vanishes, and pure graphene
is a `semi-metal'. By applying a bias voltage, the Fermi energy can move away from this symmetric point, and a circular Fermi surface develops,
as illustrated in Fig.~\ref{fig:ssgraphene}. 
The Fermi surface is electron-like for one sign of the bias, and hole-like for the other sign.
This change from electron to hole character as a function of bias voltage constitutes the quantum phase transition in graphene.
As we will see below, with regard to its dynamic properties near zero bias, graphene behaves in almost all respects like a canonical quantum critical system.

The field theory for graphene involves fermionic degrees of freedom. Representing the electronic orbitals near one of the Dirac points by
the two-component fermionic spinor $\Psi_a$, where $a$ is a sublattice index (we suppress spin and `valley' indices), we have the effective
electronic action
\begin{eqnarray}
\mathcal{S}_\Psi &=& \int d^2 r \int d \tau\,  \Psi^\dagger_a \left[ (\partial_\tau + 
i A_\tau - \mu) \delta_{ab} + i v_F \tau^x_{ab} \partial_x
+ i v_F \tau^y_{ab} \partial_y \right] \Psi_b \nonumber \\
&~&~~~~~~~~~~~~+ \frac{1}{2g^2} \int \frac{d^2 q}{4 \pi^2} \int d\tau\, \frac{q}{2\pi} \left| A_\tau ({\bf q},\tau) \right|^2, 
\label{eq:ssgraph}
\end{eqnarray}
where $\tau^i_{ab}$ are Pauli matrices in the sublattice space, $\mu$ is the chemical potential, $v_F$ is the Fermi velocity,
and $A_\tau$ is the scalar potential mediating the Coulomb interaction
with coupling $g^2 = e^2/\epsilon$ ($\epsilon$ is a dielectric constant). 
This theory undergoes a quantum phase transition as a function of $\mu$, at $\mu=0$,
similar in many ways to that of $\mathcal{S}_{LG}$ as a function of $s$. The interaction between the fermionic excitations here
has coupling $g^2$, which is the analog of the non-linearity $u$ in $\mathcal{S}_{LG}$. However, while $u$ scaled to a fixed point
non-zero fixed point value under the renormalization group flow, $g$ flows logarithmically slowly to zero. For many purposes, it is safe
to ignore this flow, and to set $g$ equal to fixed value; this value is characterized by the dimensionless `fine structure constant'
$\alpha = g^2/(\hbar v_F)$ which is of order unity in graphene.

\subsection{Spin density waves}
\label{sec:sssdw}

Finally, we consider the onset of N\'eel order, as in Section~\ref{sec:sslgw}, but in a metal rather than an insulator. It is conventional
to refer to such metallic N\'eel states as have spin density wave (SDW) order. Our discussion here is motivated by application to the cuprate
superconductors: there is good evidence \cite{ssabba,ssgross} that the transition we describe below is present in the electron-doped cuprates, and proposals of its
application to the hole-doped cuprates will be discussed in Section~\ref{sec:sstc}.

\begin{figure}
\centering
 \includegraphics[width=\linewidth]{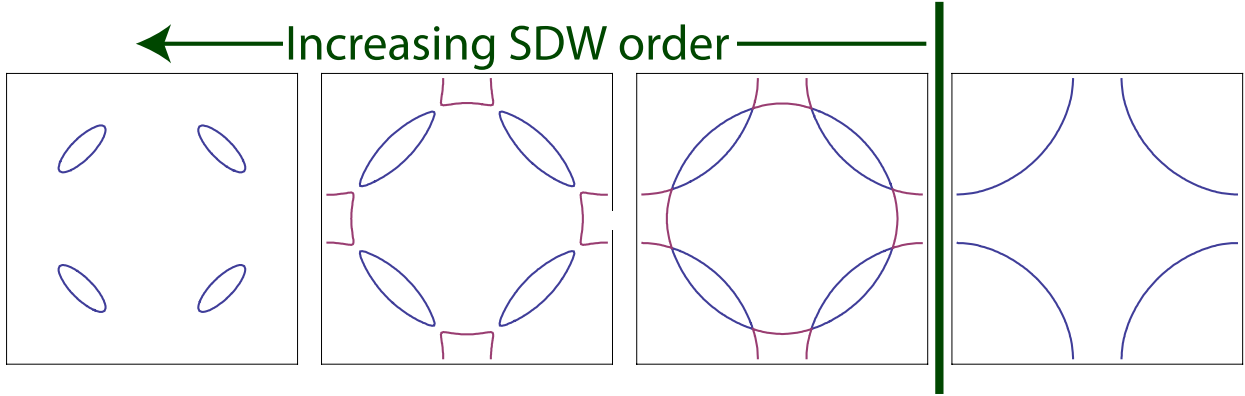}
 \caption{Evolution of the Fermi surface of the hole doped cuprates in a conventional SDW theory \cite{ssscs}
 as a function of the magnitude of the SDW order $|\varphi^a|$. The right panel
is the large Fermi surface state with no SDW order, with states contiguous to ${\bf k}=0$ occupied by electrons. 
The onset of SDW order induces the formation of electron  
(red) and hole (blue) pockets. With further increase of $|\varphi^a|$, the electron pockets disappear and only 
hole pockets remain (the converse happens in the last step for the electron-doped cuprates.}
\label{fig:sssdw}
\end{figure}

We begin with the band structure describing the cuprates in the over-doped region, well away from the Mott insulator. Here the electrons
$c_{i \alpha}$ are described by the Hamiltonian
\begin{displaymath}
H_c = - \sum_{i<j} t_{ij} c^{\dagger}_{i \alpha} c_{i \alpha} \equiv \sum_{{\bf k}} \varepsilon_{{\bf k}} c^{\dagger}_{{\bf k}\alpha} c_{{\bf k}\alpha}
\end{displaymath}
with $t_{ij}$ non-zero for first, second and third neighbors on the square lattice. This leads to Fermi surface shown in the
right-most panel of Fig.~\ref{fig:sssdw}. As in Section~\ref{sec:ssgraphene}, we will consider topological changes to this
Fermi surface, but is induced here by a conventional SDW (N\'eel) order parameter $\varphi^a$. From the structure of the ordering
in Fig.~\ref{fig:ssdimer} we see that $\varphi^a$ carries momentum ${\bf K}= (\pi,\pi)$, and so will transfer momentum ${\bf K}$
while scattering the $c_{\alpha}$ fermions; this leads to the coupling 
\begin{displaymath}
H_{\rm sdw} = {\varphi}^a \, \sum_{{\bf k}, \alpha, \beta} c_{{\bf k}, \alpha}^\dagger{\sigma}^a_{\alpha\beta}
c_{{\bf k} + {\bf K}, \beta}.
\end{displaymath}

We can now follow the evolution of the Fermi surface under the onset of SDW order by diagonalizing $H_c + H_{\rm sdw}$ for constant
$\varphi^a$: the results of this \cite{ssscs} 
are shown in Fig.~\ref{fig:sssdw}. The second panel from the right shows the Fermi surface obtained
by translating the original Fermi surface by ${\bf K}$, and the remaining panels show the consequences of mixing between the states
at momentum ${\bf k}$ and ${\bf k} + {\bf K}$.

We are now interested in the nature of the quantum critical point at the onset of SDW order, indicated by the vertical
line in Fig.~\ref{fig:sssdw}. Note that this transition combines the features of Sections~\ref{sec:sslgw}
and~\ref{sec:ssgraphene}: it has an order parameter, as in the coupled dimer antiferromagnet, and it has a topological
change in the Fermi surface, as in graphene. Thus a complete theory is just the combination of these two features:
$\mathcal{S}_{LG} + H_c + H_{\rm sdw}$. A common assertion \cite{ssmaki,sshertz} is that we can pay less attention to the Fermi surface
change by simply integrating out the $c_{\alpha}$ fermions and working with the resulting modified action for $\varphi^a$.
Right at the quantum critical point,
the SDW fluctuations, $\varphi^a$ 
connect points on the large Fermi surface, and so can decay into a large density of states
of particle-hole excitations. The damping induced by this particle-hole continuum modifies the effective action
for $\varphi^a$ from Eq.~(\ref{eq:ssslg}) by adding a relevant dissipative term: \cite{ssmaki,sshertz}
\begin{equation}
\mathcal{S}_H = \mathcal{S}_{LG} + \int \frac{d^2 k}{4 \pi^2} \int \frac{d \omega}{2 \pi} |\omega| |\varphi^a (k, \omega)|^2
\label{eq:sshertz}
\end{equation}
More recent analyses \cite{ssabanov} have indicated that this procedure is likely not correct in two spatial dimensions,
and that the Fermi surface change has to be treated in a more fundamental manner.

\section{Finite temperature crossovers}
\label{sec:sscross}
The previous section has described four model systems at $T=0$: we examined the change in the nature of the ground
state as a function of some tuning parameter, and motivated a quantum field theory which describes the low energy excitations
on both sides of the quantum critical point.

\begin{figure}
\centering
 \includegraphics[width=4.5in]{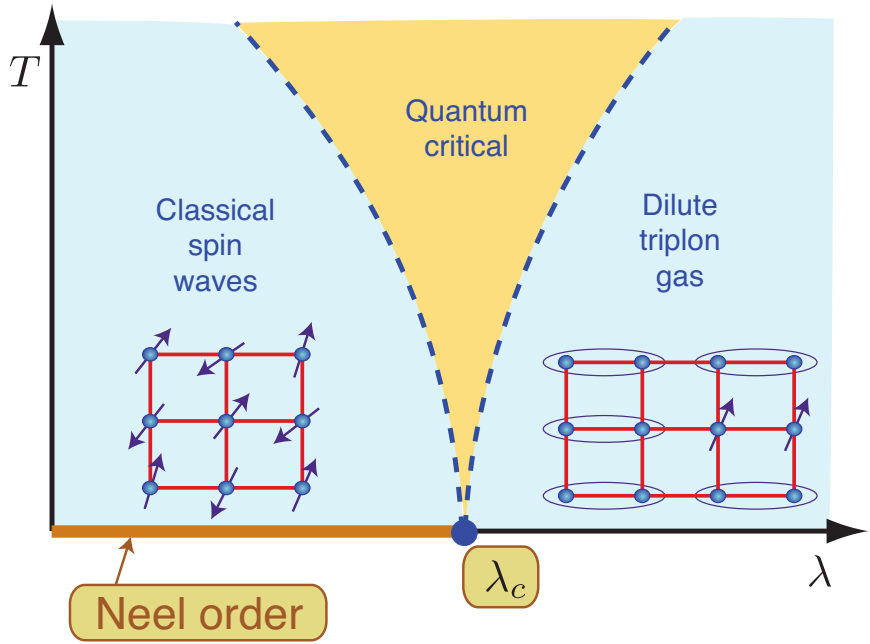}
 \caption{Finite temperature crossovers of the coupled dimer antiferromagnet in Fig.~\ref{fig:ssdimer}.}
\label{fig:sscross_dimer}
\end{figure}

We now turn to the important question of the physics at non-zero temperatures. All of the models share some common features,
which we will first explore for the coupled dimer antiferromagnet.  For $\lambda > \lambda_c$ (or $s>0$ in $\mathcal{S}_{LG}$),
the excitations consist of a triplet of $S=1$ particles (the `triplons'), which can be understood perturbatively in the large
$\lambda$ expansion as an excited $S=1$ state on a dimer, hopping between dimers (see Fig.~\ref{fig:sscross_dimer}).
The mean field theory tells us that the excitation energy of this dimer vanishes as $\sqrt{s}$ upon approaching
the quantum critical point. Fluctuations beyond mean field, described by $\mathcal{S}_{LG}$, show that the exponent
is modified to $s^{z \nu}$, where $z=1$ is the dynamic critical exponent, and $\nu$ is the correlation length exponent.
Now imagine turning on a non-zero temperature. As long as $T$ is smaller than the triplon gap, {\em i.e.\/}
$T < s^{z \nu}$, we expect a description in terms of a dilute gas of thermally excited triplon particles. This leads to the behavior
shown on the right-hand-side of Fig.~\ref{fig:sscross_dimer}, delimited by the crossover indicted by the dashed line.
Note that the crossover line approaches $T=0$ only at the quantum critical point.

Now let us look a the complementary behavior at $T>0$ on the N\'eel-ordered side of the transition, with $s<0$. 
In two spatial dimensions, 
thermal fluctuations prohibit the breaking of a non-Abelian symmetry at all $T>0$, and so spin rotation symmetry
is immediately restored. Nevertheless, there is an exponentially large spin correlation length, $\xi$, and at distances shorter
than $\xi$ we can use the ordered ground state to understand the nature of the excitations. Along with the spin-waves, we also
found the longitudinal `Higgs' mode with energy $\sqrt{-2s}$ in mean field theory. Thus, just as was this case for $s>0$, we
expect this spin-wave+Higgs picture to apply at all temperatures lower than the natural energy scale; {\em i.e.\/} 
for $T<(-s)^{z\nu}$. This leads to the crossover boundary shown on the left-hand-side
of Fig.~\ref{fig:sscross_dimer}. 

Having delineated the physics on the two sides of the transition, we are left with the crucial {\em quantum critical\/} region in the
center of Fig.~\ref{fig:sscross_dimer}. This is present for $T > |s|^{z \nu}$, {\em i.e.\/} at {\em higher\/} temperatures in the
vicinity of the quantum critical point. To the left of the quantum critical region, we have a description of the dynamics and transport
in terms of an effectively classical model of spin waves: this is the `renormalized classical' regime of Ref.~\cite{sschn}.
To the right of the quantum critical region, we again have a regime of classical dynamics, but now in terms of a Boltzmann
equation for the triplon particles. A key property of quantum critical region is that there is no description in terms
of either classical particles or classical waves at the times of order the typical relaxation time, $\tau_r$, of thermal excitations.
Instead, quantum and thermal effects are equally important, and involve the non-trivial dynamics of the fixed-point
theory describing the quantum critical point. Note that while the fixed-point theory applies only at a single point ($\lambda=\lambda_c$)
at $T=0$, its influence broadens into the quantum critical region at $T>0$. Because there is no characteristing energy scale
associated with the fixed-point theory, $k_B T$ is the only energy scale available to determine $\tau_r$ at non-zero
temperatures. Thus, in the quantum critical region
\begin{equation}
\tau_r = \mathcal{C}\frac{\hbar}{k_B T}
\label{eq:ssrelax}
\end{equation}
where $\mathcal{C}$ is a universal constant dependent only upon the universality class of the fixed point theory {\em i.e.\/}
it is universal number just like the critical exponents. This value of $\tau_r$ determines the `friction coefficients' associated
with the dissipative relaxation of spin fluctuations in the quantum critical region. It is also important for the transport
co-efficients associated with conserved quantities, and this will be discussed in Section~\ref{sec:sstrans}.

Let us now consider the similar $T>0$ crossovers for the other models of Section~\ref{sec:ssmodel}.

The N\'eel-VBS transition of Section~\ref{sec:ssdeconfine}
has crossovers very similar to those in Fig.~\ref{fig:sscross_dimer}, with one important
difference. The VBS state breaks a discrete lattice symmetry, and this symmetry remains broken for a finite range
of non-zero temperatures. Thus, within the right-hand 'triplon gas' regime of Fig.~\ref{fig:sscross_dimer}, there is a 
phase transition line at a critical temperature $T_{\rm VBS}$. The value of $T_{\rm VBS}$ vanishes very rapidly
as $s \searrow 0$, and is controlled by the non-perturbative monopole effects which were briefly noted
in Section~\ref{sec:ssdeconfine}.

\begin{figure}
\centering
 \includegraphics[width=3.8in]{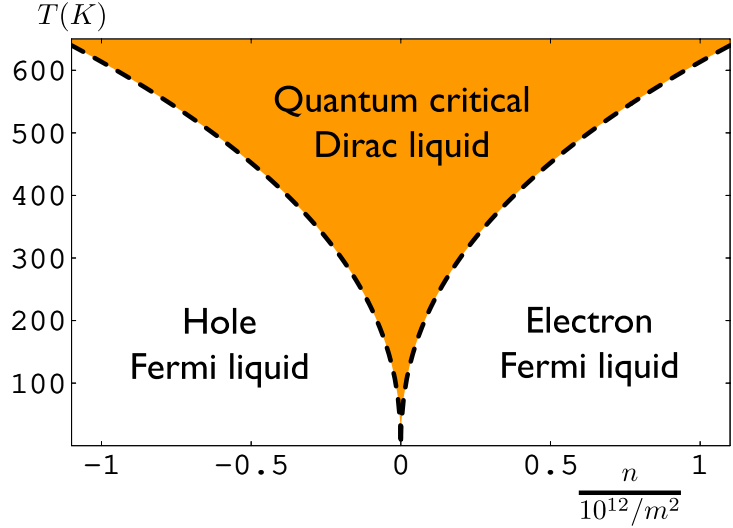}
 \caption{Finite temperature crossovers of graphene as a function of electron density $n$ (which is tuned by $\mu$ in
 Eq.~(\ref{eq:ssgraph})) and temperature, $T$. Adapted from Ref.~\onlinecite{sssheehy}.}
\label{fig:sscross_graphene}
\end{figure}
\begin{figure}
\centering
 \includegraphics[width=4.2in]{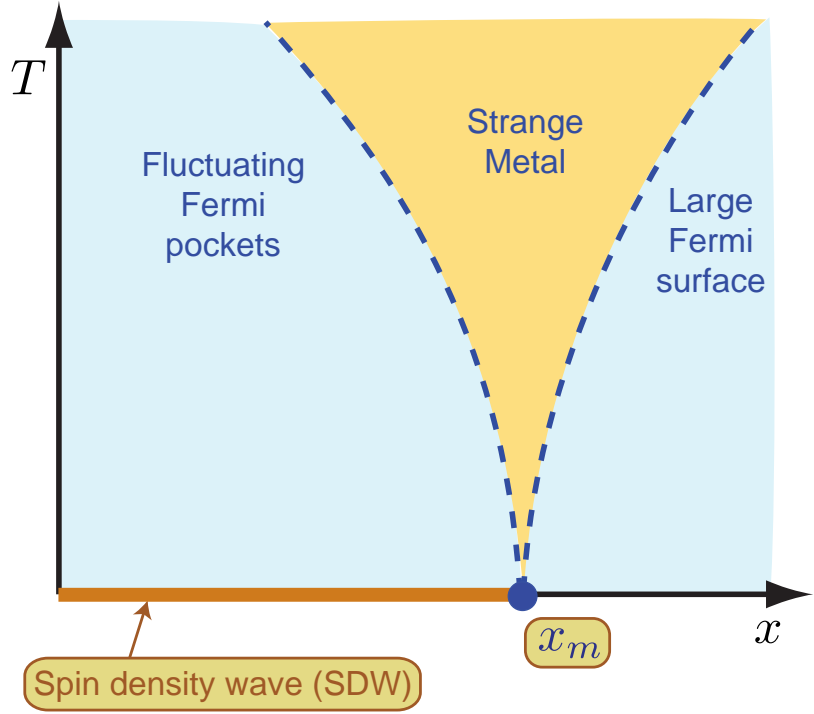}
 \caption{Finite temperature crossovers of near the SDW ordering transition of Fig.~\ref{fig:sssdw}. Here, looking to application to
 the cuprates to be discussed in Section~\ref{sec:sstc}, we have assumed that the carrier density, $x$, tunes the system across the SDW transition.}
\label{fig:sscross_sdw}
\end{figure}
For graphene, the discussion above applied to Fig.~\ref{fig:ssgraphene} leads to the crossover diagram
shown in Fig.~\ref{fig:sscross_graphene}, as noted by Sheehy and Schmalian \cite{sssheehy}.
We have the Fermi liquid regimes of the electron- and hole-like Fermi surfaces on either side of the critical point,
along with an intermediate quantum critical Dirac liquid. A new feature here is related to the logarithmic flow of the 
dimensionless `fine structure constant' $\alpha$ controlling the Coulomb interactions, which was noted in Section~\ref{sec:ssgraphene}.
In the quantum critical region, this constant takes the typical value $\alpha \sim 1/\ln (1/T)$. Consequently for the relaxation
time in Eq.~(\ref{eq:ssrelax}) we have $\mathcal{C} \sim \ln^2 (1/T)$. This time determines both the width of the
electron spectral functions, and also the transport co-efficients, as we will see in Section~\ref{sec:sstrans}.

Finally, we turn to the spin density wave transition of Section~\ref{sec:sssdw}. From the evolution in Fig.~\ref{fig:sssdw}
and the discussion above, we have the crossover phase diagram \cite{ssmillis} in Fig.~\ref{fig:sscross_sdw}.
As in Fig.~\ref{fig:sscross_dimer}, there is no transition at non-zero temperatures, and now the crossover
is between the topologically distinct Fermi surface configurations of Fig.~\ref{fig:sssdw}. The Hertz action for the
SDW fluctuations in Eq.~(\ref{eq:sshertz})
predicts \cite{ssmillis,ssabanov} logarithmic corrections to the leading scaling behavior, similar to those for graphene. 
The interplay of such SDW fluctuations with the topological change in the Fermi surface configuration is not fully understood \cite{ssvojta},
and is labeled as the `strange metal' regime in Fig.~\ref{fig:sscross_sdw}.

\section{Quantum critical transport}
\label{sec:sstrans}

We now turn to the `transport' properties in the quantum critical region: we consider the response
functions associated with any globally conserved quantity.
For the antiferromagnetic systems in Sections~\ref{sec:sslgw} and~\ref{sec:ssdeconfine},
this requires consideration of the transport of total spin, and the associated spin conductivities and diffusivities.
For graphene, we can consider charge and momentum transport. Our discussion below will also apply to the 
superfluid-insulator transition: for bosons in a periodic potential, this transition is described \cite{ssfwgf} by a field theory
closely related to that in Eq.~(\ref{eq:ssslg}). However, we will primarily use a language appropriate to charge transport
in graphene below. We will describe the properties of a generic strongly-coupled quantum critical point and mention, where
appropriate,
the changes due to the logarithmic flow of the coupling in graphene.

In traditional condensed matter physics, transport is described by identifying the low-lying
excitations of the quantum ground state, and writing down `transport equations' for the conserved charges carried by them.
Often, these excitations have a particle-like nature, such as the `triplon' particles of Fig.~\ref{fig:sscross_dimer} or the electron
or hole quasiparticles of the Fermi liquids in Fig.~\ref{fig:sscross_graphene}. In other cases, the low-lying excitations are waves, such as the spin-waves in Fig.~\ref{fig:sscross_dimer}, and their transport is described by a non-linear wave equation (such as the 
Gross-Pitaevski equation). 
However, as we have discussed in Section~\ref{sec:sscross} neither description is 
possible in the quantum critical region, because the excitations do not have a particle-like
or wave-like character. 

Despite the absence of an intuitive description of the quantum critical dynamics, we can expect that the transport
properties should have a universal character determined by the quantum field theory of the quantum critical point. In addition
to describing single excitations, this field theory also determines the $S$-matrix of these excitations by the renormalization
group fixed-point value of the couplings, and these should be sufficient to determine transport properties \cite{ssdamle}. 
The transport co-efficients, and the relaxation time to local
equilibrium, are not proportional to a mean free scattering time between the excitations, as is
the case in the Boltzmann theory of quasiparticles. Such a time would typically depend upon the interaction
strength between the particles. Rather, the system behaves like a ``perfect fluid'' in which the relaxation
time is as short as possible, and is determined universally by the absolute temperature, as indicated in Eq.~(\ref{eq:ssrelax}).
 Indeed, it was
conjectured in Ref.~\cite{ssbook} that the relaxation time in Eq.~(\ref{eq:ssrelax}) is a generic lower bound
for interacting quantum systems. Thus the non-quantum-critical regimes of all the phase diagrams in Section~\ref{sec:sscross}
have relaxation times which are all longer than Eq.~(\ref{eq:ssrelax}).

The transport co-efficients of this quantum-critical perfect fluid also do not depend upon the interaction
strength, and can be connected to the fundamental constants of nature. In particular, the electrical conductivity, $\sigma$,
is given by (in two spatial dimensions) \cite{ssdamle}
\begin{equation}
\sigma_Q = \frac{e^{\ast 2}}{h} \Phi_\sigma, \label{ssds}
\end{equation}
where $\Phi_\sigma$ is a universal dimensionless constant of order unity, and we have added the subscript $Q$ to emphasize that this is the conductivity for the case of graphene with the Fermi level at the Dirac point (for the superfluid-insulator
transition, this would correspond to bosons at integer filling) with
no impurity scattering, and at zero magnetic field. Here $e^\ast$ is the charge of the carriers: for a superfluid-insulator
transition of Cooper pairs, we have $e^\ast = 2e$, while for graphene we have $e^\ast = e$.
The renormalization group flow of the `fine structure constant' $\alpha$ of graphene to zero at asymptotically low $T$, allows
an exact computation in this case \cite{ssmarkus}: $\Phi_\sigma \approx 0.05 \ln^2 (1/T)$. For the superfluid-insulator
transition, $\Phi_\sigma$ is $T$-independent (this is the generic situation with non-zero fixed point values of the interaction \cite{sslongtime}) but it has only been computed \cite{ssbook,ssdamle} to 
leading order in expansions
in $1/N$ (where $N$ is the number of order parameter components) and in $3-d$ (where $d$ is the spatial dimensionality).
However, both expansions are neither straightforward nor rigorous, and 
require a physically motivated resummation of the
bare perturbative expansion to all orders. It would therefore be valuable to have exact solutions of quantum critical
transport where the above results can be tested, and we turn to such solutions in the next section.

In addition to charge transport, we can also consider momentum transport. This was considered in the context of applications
to the quark-gluon plasma \cite{sskss}; application of the analysis of Ref.~\cite{ssdamle} shows that the viscosity, $\eta$, is given by
\begin{equation}
\frac{\eta}{s} = \frac{\hbar}{k_B} \Phi_\eta, \label{sseta}
\end{equation}
where $s$ is the entropy density, and again $\Phi_\eta$ is a universal constant of order unity. 
The value of $\Phi_\eta$ has recently been computed \cite{ssgrapheneperfect} for graphene, and again has a logarithmic 
$T$ dependence because of the marginally irrelevant interaction: $\Phi_\eta \approx 0.008 \ln^2 (1/T)$.

We conclude this section by discussing  
some subtle aspects of the physics behind the seemingly simple result quantum-critical in Eq.~(\ref{ssds}).
For simplicity, we will consider the case of a ``relativistically'' invariant quantum critical point in 2+1 dimensions
(such as the field theories of Section~\ref{sec:sslgw} and~\ref{sec:ssdeconfine}, but marginally violated by graphene, a subtlety we ignore below).
Consider the retarded
correlation function of the charge density, $\chi (k, \omega)$, where $k = |{\bf k}|$
is the wavevector, and $\omega$ is frequency; the dynamic conductivity, $\sigma(\omega)$, is related to $\chi$ by the 
Kubo formula, 
\begin{equation}
\sigma (\omega) = \lim_{k \rightarrow 0} \frac{-i \omega}{k^2} \chi (k,
\omega).
\label{sskubo}
\end{equation}
It was argued in Ref.~\cite{ssdamle} that 
despite the absence of particle-like excitations of the critical ground state, the central characteristic of the transport
is a crossover from collisionless to collision-dominated transport. At high frequencies or low temperatures,
the limiting form for $\chi$ reduces to that at $T=0$, which is completely determined by relativistic and scale invariance 
and current conversion upto an overall constant
\begin{equation}
\chi (k, \omega)    = \frac{e^{\ast 2}}{h} K \frac{k^2}{\sqrt{v^2 k^2 - (\omega+i \eta)^2}}~~,~~\sigma(\omega) = 
\frac{e^{\ast 2}}{h} K ~~;~~~\mbox{$\hbar\omega \gg k_B T$,} 
\label{ssd2c}
\end{equation}
where $K$ is a universal number \cite{ssmpaf}.
However,  phase-randomizing collisions are intrinsically present in any strongly interacting critical point (above one
spatial dimension)
and these lead to relaxation of perturbations to local equilibrium and the consequent emergence
of hydrodynamic behavior. So at low frequencies, we have instead an Einstein relation which determines the conductivity with
\begin{equation}
\chi (k, \omega)  =  e^{\ast 2} \chi_c \frac{Dk^2}{Dk^2 - i \omega}~~,~~\sigma(\omega) = e^{\ast 2} \chi_c D = 
\frac{e^{\ast 2}}{h} \Theta_1\Theta_2~~;~~~\mbox{$\hbar\omega \ll k_B T$,} 
\label{ssd2h}
\end{equation}
where $\chi_c$ is the compressibility and $D$ is the charge diffusion constant. Quantum critical scaling arguments show that
the latter quantities obey
\begin{equation}
\chi_c = \Theta_1 \frac{k_B T}{h^2 v^2}~~,~~ D = \Theta_2 \frac{h v^2}{k_B T}, 
\end{equation}
where $\Theta_{1,2}$ are universal numbers.
A large number of papers in the literature, 
particularly those on critical points in quantum Hall systems, have used the collisionless method of Eq.~(\ref{ssd2c}) to 
compute the conductivity.
However, the correct d.c. limit is given by Eq.~(\ref{ssd2h}),
and the universal constant in Eq.~(\ref{ssds}) is given by $\Phi_{\sigma} = \Theta_1 \Theta_2$.
Given the distinct physical interpretation of the collisionless and collision-dominated regimes,
we expect that $K \neq \Theta_1 \Theta_2$.
This has been shown in a resummed perturbation expansion for a number of 
quantum critical points \cite{ssbook}.

\section{Exact results for quantum critical transport}
\label{sec:ssexact}

The results of Section~\ref{sec:sstrans} were obtained by using physical arguments
to motivate resummations of perturbative expansions. Here we shall support the ad hoc
assumptions behind these results by examining an exactly solvable model of quantum critical transport.

The solvable model may be viewed as a generalization of the gauge theory in Eq.~(\ref{eq:ssSz}) to the maximal
possible supersymmetry. In 2+1 dimensions, this is known as $\mathcal{N}=8$ supersymmetry. Such a  theory with
the U(1) gauge group is free, and so we consider the non-Abelian Yang-Millis theory with a SU($N$) gauge group. The resulting
supersymmetric Yang-Mills (SYM) theory has only one coupling constant, which is the analog of the electric charge $g$
in Eq.~(\ref{eq:ssSz}). The matter content is naturally more complicated than the complex scalar $z_\alpha$ in 
Eq.~(\ref{eq:ssSz}), and also involves relativistic Dirac fermions as in Eq.~(\ref{eq:ssgraph}). However all the terms in the action
for the matter fields are also uniquely fixed by the single coupling constant $g$. Under the renormalization group, it is believed
that $g$ flows to an attractive fixed point at a non-zero coupling $g=g^\ast$; the fixed point then defines
a supersymmetric conformal field theory in 2+1 dimensions (a SCFT3), and we are interested here in the transport
properties of this SCFT3.

A remarkable recent advance has been the exact solution of this SCFT3 in the $N\rightarrow \infty$ limit
using the AdS/CFT correspondence \cite{ssimsy}.
The solution proceeds by a dual formulation as a four-dimensional supergravity theory on a spacetime with uniform negative 
curvature: anti-de Sitter space, or AdS$_4$. Remarkably, the solution is also easily extended to non-zero temperatures,
and allows direct computation of the correlators of conserved charges in real time. 
At $T>0$ a black hole appears in the gravity, resulting
in an AdS-Schwarzschild spacetime, and $T$ is also the Hawking temperature of the black hole; the real time solutions
also extend to $T>0$.

\begin{figure}
\centering
 \includegraphics[width=4.0in]{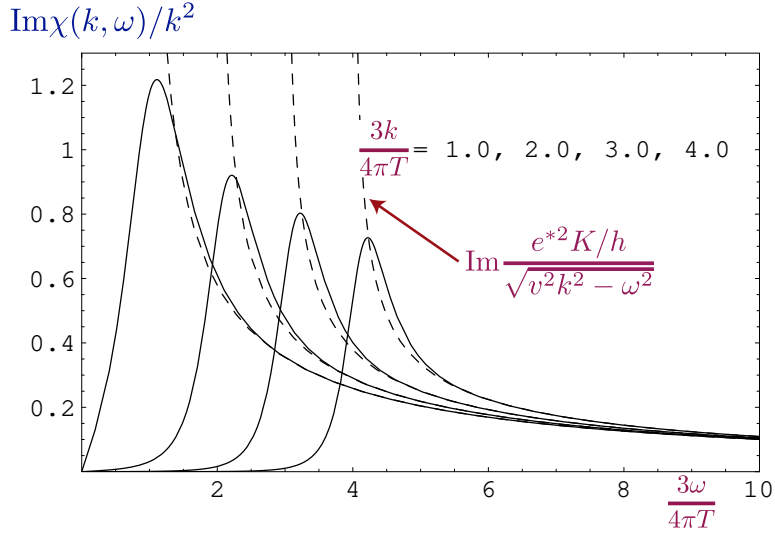}
 \caption{Spectral weight of the density correlation function of the SCFT3 with $\mathcal{N}=8$
 supersymmetry
 in the collisionless regime.}
\label{fig:sschi_collisionless}
\end{figure}

\begin{figure}
\centering
 \includegraphics[width=4.0in]{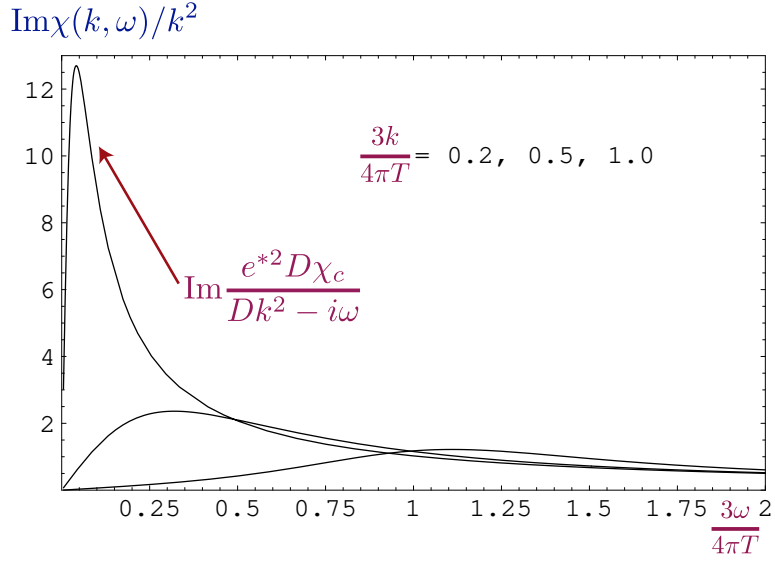}
 \caption{As in Fig.~\ref{fig:sschi_collisionless}, but for the collision-dominated regime.}
\label{fig:sschi_diff}
\end{figure}

The results of a full computation \cite{ssm2cft} of the density correlation function, $\chi (k, \omega)$ are shown
in Fig.~\ref{fig:sschi_collisionless} and~\ref{fig:sschi_diff}.
The most important feature of these results is that the expected limiting forms
in the collisionless (Eq.~(\ref{ssd2c})) and collision-dominated (Eq.~(\ref{ssd2h}) are obeyed.
Thus the results do display the collisionless to collision-dominated crossover at a frequency of order
$k_B T/\hbar$, as was postulated in Section~\ref{sec:sstrans}.

An additional important feature of the solution is apparent upon describing the full structure of 
both the density and current correlations. Using spacetime indices ($\mu,\nu=t,x,y$) we can represent these
as the tensor $\chi_{\mu\nu} ({\bf k}, \omega)$, where the previously considered $\chi \equiv \chi_{tt}$.
At $T>0$, we do not expect $\chi_{\mu\nu}$ to be relativistically covariant, and so can only constrain it by
spatial isotropy and density conservation. Introducing a spacetime momentum $p_\mu = (\omega, {\bf k})$,
and setting the velocity $v=1$,  these two constraints lead to the most general form
\begin{equation}
\chi_{\mu\nu} ({\bf k}, \omega) = \frac{e^{\ast 2}}{h} \sqrt{p^2} \Bigl( P^T_{\mu\nu}\, K^T (k,\omega)
  + P^L_{\mu\nu}\, K^L (k,\omega) \Bigr)
\label{eq:sscmn}
\end{equation}
where $p^2 = \eta^{\mu\nu} p_\mu p_\nu$ with $\eta_{\mu \nu} = \mbox{diag}(-1,1,1)$,
and $P^T_{\mu\nu}$ and $P^L_{\mu\nu}$ are
orthogonal projectors defined by
\begin{equation}
P^T_{00} = P^T_{0i} = P^T_{i0}=0~~,~~P^T_{ij} = \delta_{ij} -
\frac{k_i k_j}{k^2}~~,~~P^L_{\mu\nu} =
  \Big(\eta_{\mu\nu} - \frac{p_\mu p_\nu}{p^2}\Big) - P^T_{\mu\nu},
\end{equation}
with the indices $i,j$ running over the 2 spatial components. The two functions $K^{T,L} (k, \omega)$ define all the
correlators of the density and the current, and the results in Eqs.~(\ref{ssd2h}) and (\ref{ssd2c}) are obtained by
taking suitable limits of these functions. We will also need below the general identity 
\begin{equation}
K^T (0,\omega) = K^L (0,\omega),
\label{eq:sslt}
\end{equation}
which follows from the analyticity of the $T>0$ current correlations at ${\bf k } = 0$.

The relations of the previous paragraph are completely general and apply to any theory.
Specializing to the AdS-Schwarzschild solution of SYM3, the results were found to obey a simple
and remarkable identity \cite{ssm2cft}:
\begin{equation}
K^L (k, \omega) K^T (k, \omega) = \mathcal{K}^2
\label{eq:sssdual}
\end{equation}
where $\mathcal{K}$ is a known pure number, independent of $\omega$ and $k$. It was also shown that such a relation
applies to any theory which is equated to classical gravity on AdS$_4$, and is a consequence of the electromagnetic self-duality of its four-dimensional Maxwell sector. The combination of Eqs.~(\ref{eq:sslt})
and (\ref{eq:sssdual}) fully determines the $\chi_{\mu\nu}$ correlators at ${\bf k} = 0$: we find 
$K^L (0, \omega) = K^T (0, \omega) = \mathcal{K}$, from which it follows that the ${\bf k}=0$ conductivity
is frequency independent and that $\Phi_\sigma = \Theta_1 \Theta_2 = K= \mathcal{K}$. These last features are believed
to be special to theories which are equivalent to classical gravity, and not hold more generally.

We can obtain further insight into the interpretation of Eq.~(\ref{eq:sssdual}) by considering the field theory of the
superfluid-insulator transition of lattice bosons at integer filling. As we noted earliear, this is given by the 
field theory in Eq.~(\ref{eq:ssslg}) with the field $\varphi^a$ having 2 components. It is known that this 2-component
theory of relativistic bosons is equivalent to a dual relativistic theory, $\widetilde{\mathcal{S}}$ of vortices, under the well-known `particle-vortex'
duality \cite{ssdh}. Ref.~\cite{ssm2cft} considered the action of this particle-vortex duality on the correlation functions
in Eq.~(\ref{eq:sscmn}), and found the following interesting relations:
\begin{equation}
K^L (k, \omega) \widetilde{K}^T (k, \omega) = 1~~~,~~~K^T (k, \omega) \widetilde{K}^L (k, \omega) = 1
\label{eq:ssdual}
\end{equation}
where $\widetilde{K}^{L,T}$ determine the vortex current correlations in $\widetilde{\mathcal{S}}$ as in Eq.~(\ref{eq:sscmn}). 
Unlike Eq.~(\ref{eq:sssdual}),
Eq.~(\ref{eq:ssdual}) does {\em not} fully determine the correlation functions at ${\bf k} = 0$: it only serves
to reduce the 4 unknown functions $K^{L,T}$, $\widetilde{K}^{L,T}$ to 2 unknown functions. The key property here
is that while the theories $\mathcal{S}_{LG}$ and $\widetilde{\mathcal{S}}$ are dual to each other, they are not equivalent, and the theory $\mathcal{S}_{LG}$ is not self-dual.

We now see that Eq.~(\ref{eq:sssdual}) implies that the classical gravity theory of SYM3 is self-dual under
an analog of particle-vortex duality \cite{ssm2cft}. It is not expected that this self-duality will hold
when quantum gravity corrections are included; equivalently, the SYM3 at finite $N$ is expected
to have a frequency dependence in its conductivity at ${\bf k} = 0$. If we
apply the AdS/CFT correspondence to the superfluid-insulator transition, and approximate the latter theory
by classical gravity on AdS$_4$,  we immediately obtain the self-dual prediction for the conductivity, $\Phi_\sigma = 1$.
This value is not far from that observed in numerous experiments, and we propose here that the AdS/CFT correspondence
offers a rationale for understanding such observations.

\section{Hydrodynamic theory}
\label{sec:sshydro}

The successful comparison between the general considerations of Section~\ref{sec:sstrans}, and the exact solution
using the AdS/CFT correspondence in Section~\ref{sec:ssexact}, emboldens us to seek a more general theory of
low frequency ($\hbar \omega \ll k_B T$) transport in the quantum critical regime. We will again present our results
for the special case of a relativistic quantum critical point in 2+1 dimensions (a CFT3), but it is clear that similar considerations
apply to a wider class of systems. Thus we can envisage applications to the superfluid-insulator transition, and have presented scenarios under which
such a framework can be used to interpret measurements of the Nernst effect in the cuprates \cite{ssnernst}. 
We have also described a separate set of applications to graphene \cite{ssmarkus}: 
while graphene is strictly not a CFT3, the Dirac spectrum of electrons
leads to many similar results, especially in the inelastic collision-dominated regime associated with the quantum critical region.
These results on graphene are reviewed in a separate paper \cite{ssgraphrev}, where explicit microscopic computations are also discussed.

Our idea is to relax the restricted set of conditions under which the results of Section~\ref{sec:sstrans} were obtained.
We will work within the quantum critical regimes of the phase diagrams of Section~\ref{sec:sscross} but now allow
a variety of additional perturbations. First, we will move away from the particle-hole symmetric case, allow a finite density
of carriers. For graphene, this means that $\mu$ is no longer pinned at zero; for the antiferromagnets, we can apply
an external magnetic field; for the superfluid-insulator transition, the number density need not be commensurate
with the underlying lattice. For charged systems, such as the superfluid-insulator transition or graphene, we allow
application of an external magnetic field. Finally, we also allow a small density of impurities which can act as a sink
of the conserved total momentum of the CFT3. In all cases, the energy scale associated with these perturbations 
is assumed to be smaller than the dominant energy scale of the quantum critical region, which is $k_B T$.
The results presented below were obtained in two separate computations, associated with the methods described in 
Sections~\ref{sec:sstrans}
and~\ref{sec:ssexact}, and are described in the two subsections below.

\subsection{Relativistic magnetohydrodynamics}
\label{sec:ssmagneto}

With the picture of relaxation to local equilibrium at frequencies $\hbar \omega \ll k_B T$ developed in Ref.~\cite{ssdamle},
we postulate that the equations of relativistic magnetohydrodynamics should describe the low frequency transport.
The basic principles involved in such a hydrodynamic computation go back to the nineteenth century: conservation
of energy, momentum, and charge, and the constraint of the positivity of entropy production. Nevertheless,
the required results were not obtained until our recent work \cite{ssnernst}: the general case of a CFT3 in the presence of
a chemical potential, magnetic field, and small density of impurities is very intricate, and the guidance
provided by the dual gravity formulation was very helpful to us. In this approach, we do not have quantitative
knowledge of a few transport co-efficients, and this is complementary to our ignorance of the effective
couplings in the dual gravity theory to be discussed in Section~\ref{sec:ssdyon}.

The complete hydrodynamic analysis can be found in Ref.~\cite{ssnernst}. The analysis is intricate, but is mainly 
a straightforward adaption of the classic procedure outlined by Kadanoff and Martin \cite{sskm} to the relativistic field
theories which describe quantum critical points. We list the steps: 
\begin{enumerate}
\item Identify the conserved quantities, which are the energy-momentum tensor, $T^{\mu\nu}$, and the particle number current, $J^\mu$. 
\item Obtain the real time 
equations of motion, which express the conservation laws:
\begin{equation}
\partial_\nu T^{\mu\nu} = F^{\mu\nu}J_\nu~~~,~~~\partial_\mu J^\mu = 0;
\end{equation}
here $F^{\mu\nu}$ represents the externally applied electric and magnetic fields which can change the net
momentum or energy of the system, and we have not written a term describing momentum relaxation
by impurities. 
\item Identify the state variables which characterize the local thermodynamic state---we choose these to be the density, $\rho$, the temperature $T$, and an average velocity $u^\mu$. 
\item Express $T^{\mu\nu}$ and $J^\mu$ in terms of the state variables and their spatial and temporal
gradients; here we use the properties of the observables under a boost by the velocity $u^\mu$, and thermodynamic quantities like the energy density, $\varepsilon$, and the pressure,
$P$, which are determined from $T$ and $\rho$ by the equation of state of the CFT. We also introduce transport 
co-efficients associated with the gradient terms. 
\item Express the equations of motion in terms
of the state variables, and ensure that the entropy production rate is positive \cite{ssll}. This is a key step which ensures
relaxation to local equilibrium, and leads to important constraints on the transport co-efficients. In $d=2$, it was found that situations with the velocity $u^\mu$ spacetime independent are characterized
by only a {\em single\/} independent transport co-efficient \cite{ssnernst}. This we choose to be the longitudinal conductivity
at $B=0$.  
\item Solve the initial value problem for the state variables using the linearized equations
of motion. 
\item Finally, translate this solution to the linear response functions, as described in Ref.~\cite{sskm}.
\end{enumerate}

\subsection{Dyonic black hole}
\label{sec:ssdyon}

Given the success of the AdS/CFT correspondence for the specific supersymmetric model in Section~\ref{sec:ssexact},
we boldly assume a similar correspondence for a generic CFT3. We assume that each CFT3 is dual to a strongly-coupled
theory of gravity on AdS$_4$. Furthermore, given the operators associated with the perturbations away from the pure
CFT3 we want to study, we can also deduce the corresponding perturbations away from the dual gravity theory.
So far, this correspondence is purely formal and not of much practical use to us. However, we now restrict our
attention to the hydrodynamic, collision dominated regime, $\hbar \omega \ll k_B T$ of the CFT3. We would like
to know the corresponding low energy effective theory describing the quantum gravity theory on AdS$_4$. 
Here, we make the simplest possible assumption: the effective theory is just the Einstein-Maxwell theory
of general relativity and electromagnetism on AdS$_4$. As in Section~\ref{sec:ssexact}, the temperature $T$ of CFT3
corresponds to introducing a black hole on AdS$_4$ whose Hawking temperature is $T$. The chemical potential, $\mu$,
of the CFT3 corresponds to an electric charge on the black hole, and the applied magnetic field maps to a 
magnetic charge on the black hole. Such a dynoic black hole solution of the Einstein-Maxwell equations is, in fact, known:
it is the Reissner-Nordstrom black hole.

We solved the classical Einstein-Maxwell equations for linearized fluctuations about the metric of a dyonic black hole
in a space which is asymptotically AdS$_4$. The results were used to obtain correlators of a CFT3 using 
the prescriptions of the AdS/CFT mapping.
As we have noted,
we have no detailed knowledge of the strongly-coupled quantum gravity theory which is dual to the CFT3 describing
the superfluid-insulator transition in condensed matter systems, or of graphene. 
Nevertheless, given our postulate that its low
energy effective field theory essentially captured by the Einstein-Maxwell theory, we can then obtain a powerful set
of results for CFT3s. 

\subsection{Results}
\label{sec:ssresults}

In the end, we obtained complete agreement between the two independent computations in Sections~\ref{sec:ssmagneto}
and~\ref{sec:ssdyon}, after allowing for their distinct equations of state. This agreement demonstrates that
the assumption of a low energy Einstein-Maxwell effective field theory for a strongly coupled theory of quantum gravity
is equivalent to the assumption of hydrodynamic transport for $\hbar \omega \ll k_B T$ in a strongly coupled CFT3.

Finally, we turn to our explicit results for quantum critical transport with $\hbar \omega \ll k_B T$. 

First, consider adding a chemical potential, $\mu$,  to the CFT3. This will induce a non-zero
number density of carriers $\rho$. The value of $\rho$ is defined so that the total charge density associated with
$\rho$ is $e^{\ast} \rho$. Then the electrical conductivity at a frequency $\omega$ is 
\begin{equation}
\sigma (\omega) = \frac{e^{\ast 2}}{h} \Phi_\sigma + \frac{e^{\ast 2} \rho^2 v^2}{(\varepsilon + P)} \frac{1}{(- i \omega + 1/\tau_{\rm imp})}
\label{sssw}
\end{equation}
In this section, we are again using the symbol $v$ to denote the characteristic velocity of the CFT3 because we will need $c$
for the physical velocity of light below. Here $\varepsilon$ is the energy density and $P$ is the pressure of the CFT3.
We have assumed a small density of impurities which lead to a momentum relaxation time $\tau_{\rm imp}$ \cite{ssnernst,sshh}.
In general, $\Phi_\sigma$, $\rho$, $\varepsilon$, $P$, and $1/\tau_{\rm imp}$ will be functions of $\mu/k_B T$
which cannot be computed by hydrodynamic considerations alone. However, apart from $\Phi_\sigma$, these quantities
are usually amenable to direct perturbative computations in the CFT3, or by quantum Monte Carlo studies.
The physical interpretation of Eq.~(\ref{sssw}) should be evident: adding a charge density $\rho$ leads to an 
additional Drude-like contribution to the conductivity. This extra current cannot be relaxed by collisions between the 
unequal density of particle and hole excitations, and so requires an impurity relaxation mechanism to yield a finite
conductivity in the d.c. limit.

Now consider thermal transport in a CFT3 with a non-zero $\mu$. The d.c. thermal conductivity, $\kappa$, is given by
\begin{equation}
\kappa = \Phi_\sigma \left( \frac{k_B^2 T}{h} \right) \left( \frac{\varepsilon + P}{k_B T \rho}
\right)^2 , \label{sskapparho}
\end{equation}
in the absence of impurity scattering, $1/\tau_{\rm imp} \rightarrow 0$. This is a Wiedemann-Franz-like relation, connecting the thermal conductivity to the electrical conductivity
in the $\mu=0$ CFT. Note that $\kappa$ diverges as $\rho \rightarrow 0$, 
and so the thermal conductivity of the $\mu=0$ CFT is infinite.

Next, turn on a small magnetic field $B$; we assume that $B$ is small enough
that the spacing between the Landau levels is not as large as $k_B T$. The case
of large Landau level spacing is also experimentally important, but cannot be addressed
by the present analysis. Initially, consider the case $\mu=0$. In this case, the result
Eq.~(\ref{sskapparho}) for the thermal conductivity is replaced by
\begin{equation}
\kappa = \frac{1}{\Phi_\sigma} 
\left( \frac{k_B^2 T}{h} \right) \left( \frac{\varepsilon + P}{k_B T B/(hc/e^\ast)}
\right)^2 \label{sskappaB}
\end{equation}
also in the absence of impurity scattering, $1/\tau_{\rm imp} \rightarrow 0$.
This result relates $\kappa$ to the electrical {\em resistance} at criticality,
and so can be viewed as Wiedemann-Franz-like relation for the vortices.
A similar $1/B^2$ dependence of $\kappa$ appeared in the Boltzmann equation
analysis of Ref.~\cite{ssbgs}, but our more general analysis applies in a wider and distinct regime
\cite{ssmarkus}, and relates the
co-efficient to other observables.

We have obtained a full set of results for the frequency-dependent thermo-electric
response functions at non-zero $B$ and $\mu$. The results are lengthy and we refer
the reader to Ref.~\cite{ssnernst} for explicit expressions. Here we only note that the characteristic feature  \cite{ssnernst,sssean2} of these results is a new {\em hydrodynamic cyclotron resonance}.
The usual cyclotron resonance occurs at the classical cyclotron frequency, which is independent of the particle density and temperature; 
further, in a Galilean-invariant system this resonance
is not broadened by electron-electron interactions alone, and requires impurities
for non-zero damping. The situtation for our hydrodynamic resonance is very different.
It occurs in a collision-dominated regime, and its frequency depends on 
the density and temperature: the explicit expression for the resonance frequency is
\begin{equation}
\omega_c = \frac{e^\ast B \rho v^2}{c (\varepsilon + P)}.
\end{equation}
Further, the cyclotron resonance involves particle and hole excitations moving
in opposite directions, and collisions between them can damp the resonance
even in the absence of impurities. Our expression for this intrinsic damping frequency is \cite{ssnernst,sssean2}
\begin{equation}
\gamma = \frac{e^{\ast 2}}{h} \Phi_\sigma \frac{ B^2 v^2}{c^2 (\varepsilon + P)},
\end{equation}
relating it to the quantum-critical conductivity as a measure of collisions between 
counter-propagating particles and holes.
We refer the reader to a separate discussion \cite{ssmarkus} of the experimental conditions under
which this hydrodynamic cyclotron resonance may be observed.

\section{The cuprate superconductors}
\label{sec:sstc}

We close this article by mentioning application to the cuprate superconductors. The phenomenology of these
materials is very involved, and they have so far resisted a {\em global\/} interpretation in terms
of the canonical quantum-critical crossover phase diagrams discussed here.
Surely, one of the important complexities is that they involve at least two order parameters: those associated
with $d$-wave superconductivity and with SDWs (both commensurate and incommensurate with the
underlying lattice). In addition there are also topological changes in the Fermi surface discussed
in Section~\ref{sec:sssdw} and Fig.~\ref{fig:sssdw}.

\begin{figure}
\centering
 \includegraphics[width=4.5in]{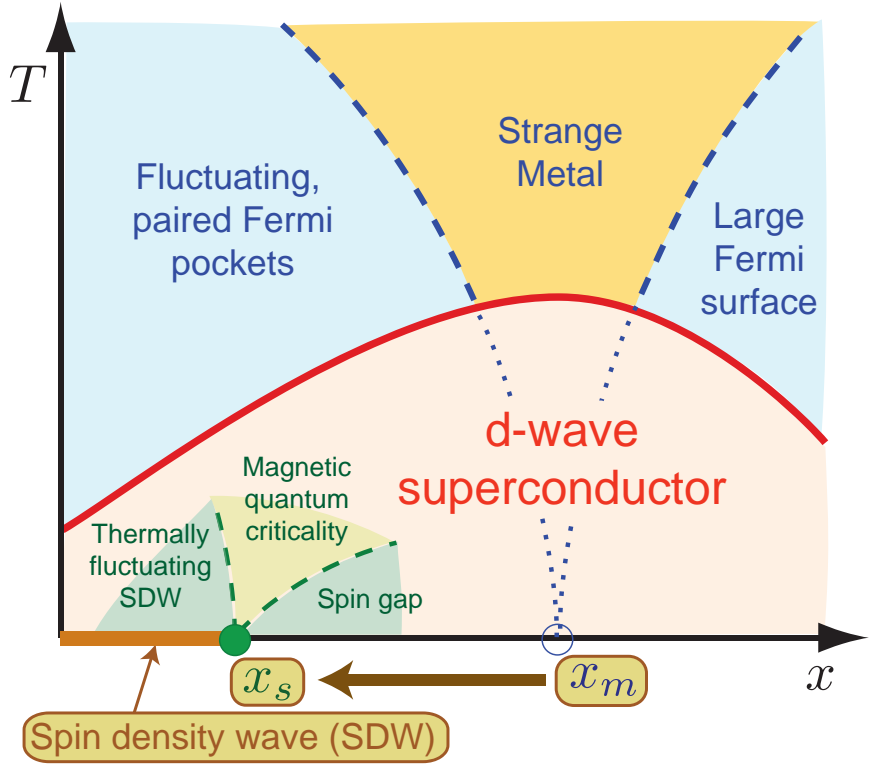}
 \caption{Influence of the onset of superconductivity on the crossovers near the SDW ordering transition in Fig.~\ref{fig:sscross_sdw}.}
\label{fig:sscross_dsc}
\end{figure}
Here we mention recent ideas \cite{ssgalitski,ssmoon,ssperspective,ssm2s} which attempt a synthesis based
upon a combination of the crossover phase diagrams we have discussed here.
The ideas build upon the results of a number of recent experiments \cite{ssdoiron,sslouis,sssuchitra,ssnlsco1,sshelm} 
which have explored
the phase diagram of both the hole- and electron-doped cuprates in a magnetic field applied
perpendicular to the layers. We do not wish to enter into the intricate details of the structure
of the phase diagram in the field-doping plane, and refer the reader to recent reviews \cite{ssperspective,ssm2s}.
We will limit ourselves to the assertion that the experiments support the proposal \cite{ssnlsco1} that the underlying
metallic state of the cuprates undergoes an SDW transition as in Fig.~\ref{fig:sscross_sdw} along with the topological change of
the Fermi surface, which was illustrated in Fig.~\ref{fig:sssdw} for the simplest case of commensurate ordering. 
Recent computations \cite{sshackl} and observations \cite{sslouisnernst} of thermoelectric effects have provided
additional support for this proposal.

The main ingredient missing from Fig.~\ref{fig:sscross_sdw} is $d$-wave superconductivity. It is known that 
the SDW fluctuations can induce $d$-wave pairing for the Fermi surface configuration shown in Fig.~\ref{fig:sssdw}.
So let us superimpose a superconducting region on Fig.~\ref{fig:sscross_sdw}, and this leads us to Fig.~\ref{fig:sscross_dsc}.
Arguments that this superconducting region will have the dome shape around the quantum critical point
were made in Ref.~\cite{ssmoon}. Here we wish to draw attention to a crucial new feature shown in Fig.~\ref{fig:sscross_dsc}.
The onset of superconductivity {\em shifts\/} the spin ordering quantum critical point from its apparent
position in the underlying metal at $x=x_m$, to its actual position in the superconductor at $x=x_s$. This shift
is indicated by the large arrow in Fig.~\ref{fig:sscross_dsc}. We have argued \cite{ssmoon,ssperspective,ssm2s} that this
shift helps resolve a number of long-standing puzzles in understanding the experimental observations in the cuprates
in terms of the quantum-critical crossover phase diagrams.

What is the physical origin of the shift in the position of the quantum critical point? In phenomenological terms, this shift can be 
understood as a {\em competition\/} between the SDW and superconducting order parameters \cite{ssdemler}.
Once there is an onset of superconductivity, it repels the incipient SDW ordering, and shifts it away from the superconducting
region to lower carrier concentration. A microscopic theory has also been presented in Ref.~\cite{ssmoon}: it requires special attention
to the physics of the Fermi surface, and the fact that SDW ordering and superconductivity are `eating up' the same portions
of the Fermi surface. Such phenomena are especially pronounced near the Mott insulating state at zero doping.

Once the shift has occurred, the crossovers near the spin-ordering transition within 
the superconductor are especially simple. There are no longer any Fermi surfaces to contend with,
and so we can largely focus on the fluctuations of the order parameter. The latter reduce \cite{ssmetlitski} to those presented
in Fig.~\ref{fig:sscross_dimer}, with critical theories like those presented in Sections~\ref{sec:sslgw}
and~\ref{sec:ssdeconfine}. 

So the bottom line is that the proposed cuprate phase diagram in Fig.~\ref{fig:sscross_dsc} is a combination of the crossovers
in Fig.~\ref{fig:sscross_dimer} and~\ref{fig:sscross_sdw}, along with the all-important shift in the position of the quantum
critical point.

It is now interesting to examine the subtle evolution in the physics as the temperature is lowered for $x_s < x < x_m$.
Above $T_c$, we have a metallic state with fluctuating SDW order whose strength increases as $T$ is lowered.
Once we cross $T_c$, the growth in SDW ordering is arrested (because of the competition with superconductivity);
the system is now to the {\em right\/} of the SDW ordering quantum critical point, and starts to recover aspects of the 
physics of the large Fermi surface. However, it has been argued \cite{ssmetlitski} that forms of translational
symmetry breaking, such as VBS or nematic ordering may appear as $T \rightarrow 0$ for $x_s < x <x_m$.
In this case, the left crossover line emerging from $x_m$ in Fig.~\ref{fig:sscross_dsc} would become an actual
thermal phase transition at which lattice symmetry is restored.

\acknowledgements
 
This research was supported by the National Science Foundation under grant DMR-0757145), 
 by the FQXi foundation, and by a MURI grant from AFOSR.

\end{document}